\documentclass[aps,prstab,twocolumn,superscriptaddress]{revtex4-1}
\usepackage{amssymb}
\usepackage{amssymb}
\usepackage{amssymb}
\usepackage{graphicx}
\usepackage{graphics}
\usepackage{amsmath}
\usepackage{placeins}
\usepackage{hyperref}
\usepackage[dvipsnames]{xcolor}

\usepackage[normalem]{ulem}

\begin{document}
	
	\title{Systematic benchmarking of planar nitrogen-incorporated ultrananocrystalline diamond field emission electron source: rf conditioning and beam spatio-temporal characteristics}
	
	\author{\firstname{Jiahang} \surname{Shao}}
	\email{jshao@anl.gov}
	\affiliation{Argonne National Laboratory, Lemont, IL 60439, USA}
	\author{\firstname{Mitchell} \surname{Schneider}}\thanks{These authors contributed equally to the work}
	\affiliation{Michigan State University, East Lansing, MI 48824, USA}
	\author{\firstname{Gongxiaohui} \surname{Chen}}\thanks{These authors contributed equally to the work}
	\affiliation{Illinois Institute of Technology, Chicago, IL 60616, USA}
	\author{\firstname{Tanvi} \surname{Nikhar}}
	\affiliation{Michigan State University, East Lansing, MI 48824, USA}
	\author{\firstname{Kiran Kumar} \surname{Kovi}}
	\affiliation{Euclid Techlabs LLC, Bolingbrook, IL 60440, USA}
	\author{\\\firstname{Linda} \surname{Spentzouris}}
	\affiliation{Illinois Institute of Technology, Chicago, IL 60616, USA}
	\author{\firstname{Eric} \surname{Wisniewski}}
	\affiliation{Argonne National Laboratory, Lemont, IL 60439, USA}
	\author{\firstname{John} \surname{Power}}
	\affiliation{Argonne National Laboratory, Lemont, IL 60439, USA}
	\author{\firstname{Manoel} \surname{Conde}}
	\affiliation{Argonne National Laboratory, Lemont, IL 60439, USA}
	\author{\firstname{Wanming} \surname{Liu}}
	\affiliation{Argonne National Laboratory, Lemont, IL 60439, USA}
	\author{\firstname{Sergey V.} \surname{Baryshev}}
	\email{serbar@msu.edu}
	\affiliation{Michigan State University, East Lansing, MI 48824, USA}
	
	\date{\today}
	
	\begin{abstract}
		Planar nitrogen-incorporated ultrananocrystalline diamond, (N)UNCD, has emerged as a unique field emission source attractive for accelerator applications because of its capability to generate high charge beam and handle moderate vacuum conditions. Most importantly, (N)UNCD sources are simple to produce: conventional high aspect ratio isolated emitters are not required to be formed on the surface, and the actual emitter surface roughness is on the order of only 100~nm. Careful reliability assessment of (N)UNCD is required before it may find routine application in accelerator systems. In the present study using an L-band normal conducting single-cell rf gun, a (N)UNCD cathode has been conditioned to $\sim$42~MV/m in a well-controlled manner. It reached a maximum output charge of 15~nC corresponding to an average current of 6~mA during an emission period of 2.5~$\mu$s. Imaging of emission current revealed a large number of isolated emitters (density over 100/cm$^{2}$) distributed on the cathode, which is consistent with previous tests in dc environments. The performance metrics, the emission imaging, and the systematic study of emission properties during rf conditioning in a wide gradient range assert (N)UNCD as an enabling electron source for rf injector designs serving industrial and scientific applications. These studies also improve the fundamental knowledge of the practical conditioning procedure via better understanding of emission mechanisms.
	\end{abstract}
	
	
	\maketitle
	
	\section{Introduction}
	The field emission cathode (FEC) is a viable choice for rf injectors in many industrial and scientific applications~\cite{BrauNIMA1998,FurseyBook2007,LeemannPRST2007,LewellenPRST2005,GanterNIMA2006,XiangkunPRST2013,PiotAPL2014,SergeyAPL2014,MihalceaAPL2015,JiaqiIEEE2018}. It has several clear advantages compared to thermionic cathodes and photocathodes. Firstly, it is simple as no heating system nor laser is required to facilitate electron emission~\cite{XiangkunPRST2013,PiotAPL2014,MihalceaAPL2015}. Secondly, it has the capability to deliver high output current with maximum current densities on the order of 10$^{8}$~A/cm$^{2}$~\cite{GanterNIMA2006}. Thirdly, it has the potential to reach ultra-low emittance and energy spread via various gating methods~\cite{LewellenPRST2005,XiangkunPRST2013,XiangkunNIMA2015,JiahangPRL2016}. Altogether, the way toward realization of high performance FEC-based injectors is being paved.
	
	Maintaining high FEC performance over long term operation remains an important goal. Various advanced field emitter materials and configurations are currently under intense investigation in dc and rf environments. For high power vacuum systems, materials of interest are often carbon derived substances such as carbon nanotube (CNT)~\cite{TeoNature2005,MihalceaAPL2015,QilongJAP2016}, nanocrystalline diamond (NCD)~\cite{ZhouJAP1997,IkedaAPL2009,NemanichMB2014}, nitrogen-incorporated ultrananocrystalline diamond ((N)UNCD)~\cite{XiangkunPRST2013,SergeyAPL2014,ChubenkoACS2017}, as well as some other forms of carbon~\cite{ShangACS2009,KrivchenkoJAP2010,TeiiAPL2010}. For aggressive high oxygen content environments like thrusters, diamond and boron nitride are under intense studies~\cite{LevchenkoNatureComm2018}. Configurations of interest include Spindt-type high aspect ratio single emitters or emitter arrays, and planar nm-roughness emitters.
	
	Planar (N)UNCD is a promising material that yields high current and allows for simplicity and scalability in fabrication. Although the exact mechanism of field emission from diamond surface is still under debate, previous experiments have shown that emission originates from the $sp^2$ grain boundaries~\cite{ChatterjeeAPL2014,HarnimanCarbon2015}. The high grain boundary density of (N)UNCD material could therefore potentially lead to high current density. To date, planar (N)UNCD cathodes have been tested and shown to perform at 20-70~MV/m in normal conducting rf guns~\cite{XiangkunPRST2013,SergeyAPL2014}, at 1~MV/m under cryogenic temperatures of 2-4~K in a superconducting rf gun~\cite{UNCDSRF}, and at 1-20~MV/m in dc setups~\cite{ChubenkoACS2017}. Like any other emitters, the full operation range, from turn-on up to the breakdown field, must be explored systematically before (N)UNCD can be considered for accelerator applications that may include industrial systems operated at surface fields of 1-20~MV/m, or scientific systems operated at surface fields of $\sim$100~MV/m. Addressing the fundamental emission mechanism puzzle is also critical for successful (N)UNCD deployment.
	
	The study presented here extends planar (N)UNCD cathode high pulsed power characterization during rf conditioning. Detailed emission properties were recorded as the macroscopic $E$-field was conditioned from $\sim$8~MV/m to $\sim$42~MV/m; these included current and current density, effective emission area, field enhancement factor and microscopic electric field, turn-on field, and temporal emission stability. The metrics evolution of the emission parameters was analyzed and interpreted in the framework of a unique density-of-states structure of (N)UNCD~\cite{ChubenkoJAP2019}. In addition, the (N)UNCD emitting surface was visualized in the rf gun environment with high resolution which showed a large number of localized emitters (density over 100/cm$^{2}$) distributed across the surface. This result is consistent with results obtained in dc setups~\cite{CuiJVST2002,KockDiamond2004,ChubenkoACS2017}.
	
	This paper is organized as follows: Sec.~\ref{cha2} presents the cathode preparation; Sec.~\ref{cha3} introduces the experimental setup; Sec.~\ref{cha4} provides the detailed description of the experimental methods, procedures and results; Sec.~\ref{cha5} discusses the experimental results and presents a hypothesis behind the emission evolution during rf conditioning; and Sec.~\ref{cha6} summarizes the study and outlines future work.
	
	\section{\label{cha2}Cathode preparation}
	\subsection{Cathode assembly}
\begin{figure}[h!tbp]
		\includegraphics[width=4.5cm]{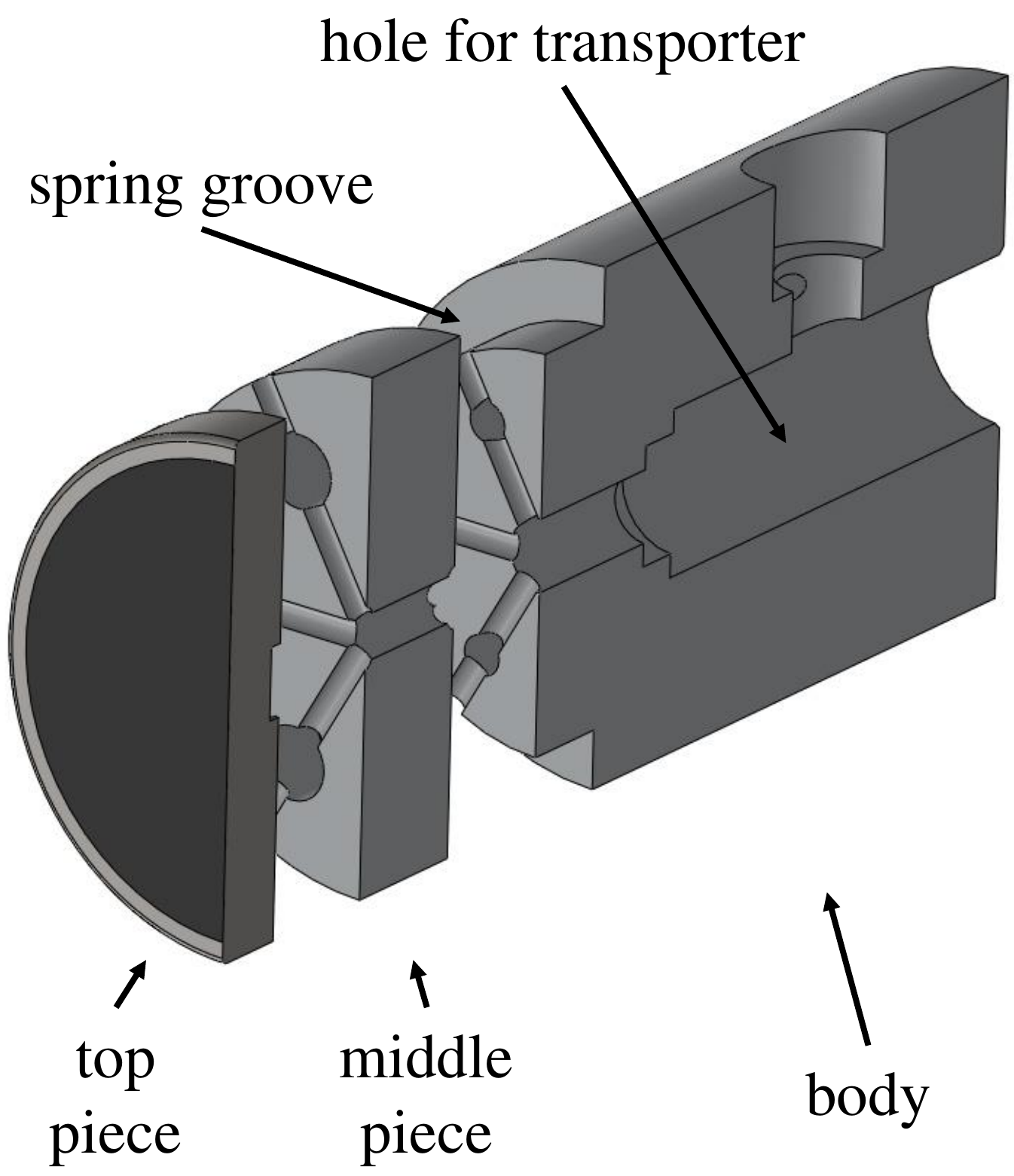}
		\caption{\label{Fig_cathode_assembly} Cutaway drawing of the cathode assembly. The dark part of the top piece represents the (N)UNCD material. The vented screws holding the parts together are not shown.}
	\end{figure}

	The cathode plug (28~mm tall and 20~mm in diameter) is designed as a three-part assembly with an aluminum body, an aluminum middle piece, and a stainless steal top piece, as illustrated in Fig.~\ref{Fig_cathode_assembly}. The design meets the installation requirements of the L-band photocathode rf gun test-stand~\cite{SergeyAPL2014,JiahangPRL2015,JiahangPRL2016,Jiahangthesis2018} and enables convenient material synthesis onto the thin top part. The parts are aligned with each other and assembled together using internal vented screws. The electrical contact between the cathode assembly and the rf gun is ensured by a spring around the cathode body. 
	
	\subsection{Cathode synthesis}
\begin{figure}[h!tbp]
		\includegraphics[width=8cm]{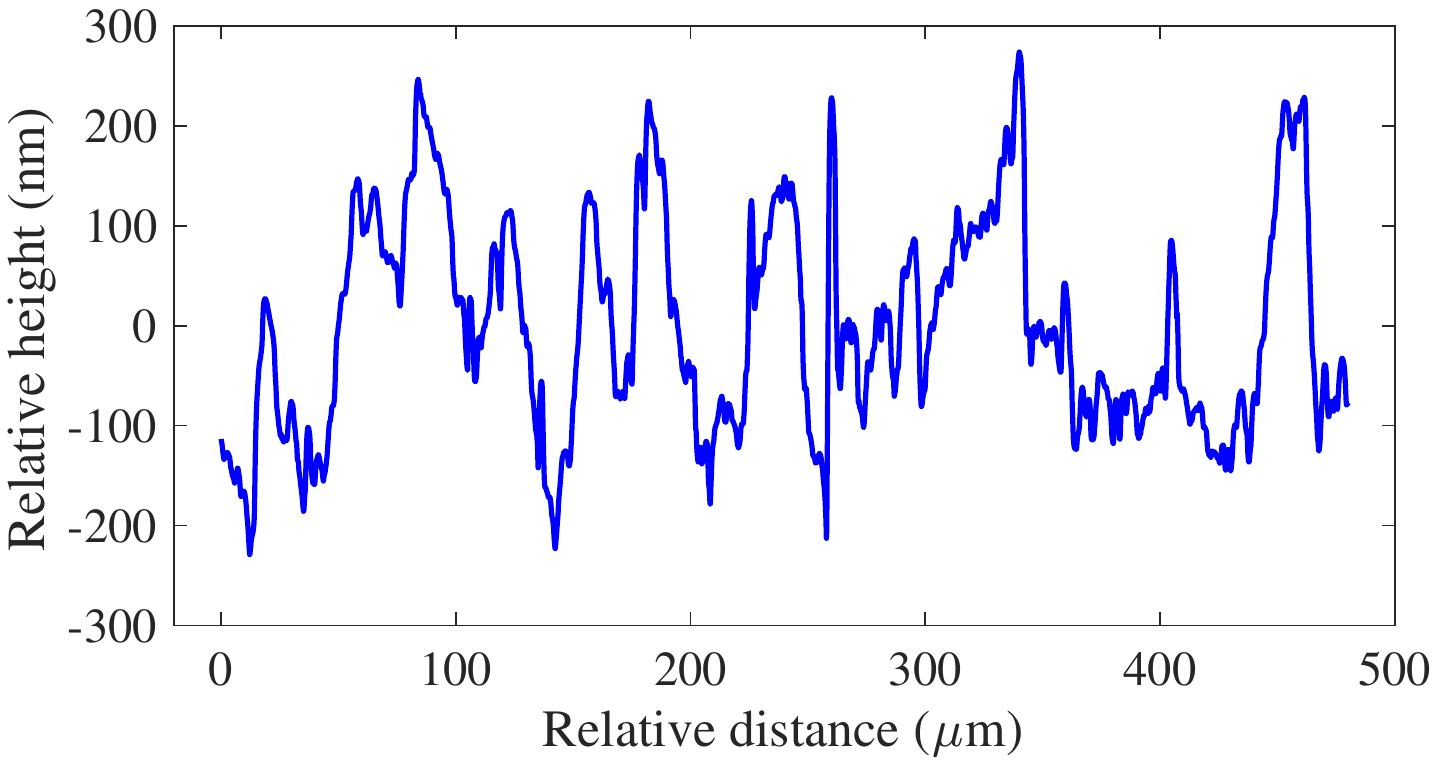}
		\caption{\label{Fig_Roughness} Surface roughness measurement of the top piece after polishing. The peak-to-peak and root-mean-square roughness is 503~nm and 108~nm, respectively.}
	\end{figure}

	The synthesis procedure of depositing the (N)UNCD material onto the top piece is described as follows. (1) The top piece was polished to 100~nm-level roughness to avoid any high aspect ratio site induced emission (See Fig.~\ref{Fig_Roughness} for the surface roughness measurement result). (2) A molybdenum buffer layer of 150~nm thickness was deposited using magnetron sputtering. (3) The top piece underwent an ultrasonic seeding procedure that made use of diamond slurry (Adamas Technologies) with a particle size of 5-10~nm. (4) The (N)UNCD material was deposited using microwave-assisted plasma chemical vapor deposition (MPCVD by Lambda Technologies Inc.) operated at 915~MHz. The growth conditions were: total pressure of 56~Torr, microwave power of 2.3~kW, individual gas flows of Ar, CH$_{4}$, and N$_{2}$ at 3~sccm, 160~sccm, and 40~sccm respectively.
	
	Growing (N)UNCD via chemical vapor deposition is a competing process of forming the diamond $sp^{3}$ phase and the graphitic $sp^{2}$ phase in the right proportion to allow  mechanical strength and adhesion, and to obtain high conductivity and emission efficiency. Here, high emission efficiency means low turn-on field (1-10~MV/m) and high field enhancement factor (100-1,000). Given that grain boundaries ($sp^2$ phase) facilitate emission, the (N)UNCD efficiency can be “tuned” through the $sp^3$-to-$sp^2$ ratio~\cite{CorriganDiamond2002,IkedaAPL2009}. Compared to our previous growth protocol executed at 850~$^\circ$C~\cite{SergeyAPL2014}, the synthesis temperature was increased by 50~$^\circ$C to induce a larger fraction of ordered graphitic $sp^2$ phase that is known to improve the emission efficiency of (N)UNCD~\cite{IkedaAPL2009}. 
	
\begin{figure}[h!tbp]
		\includegraphics[width=8cm]{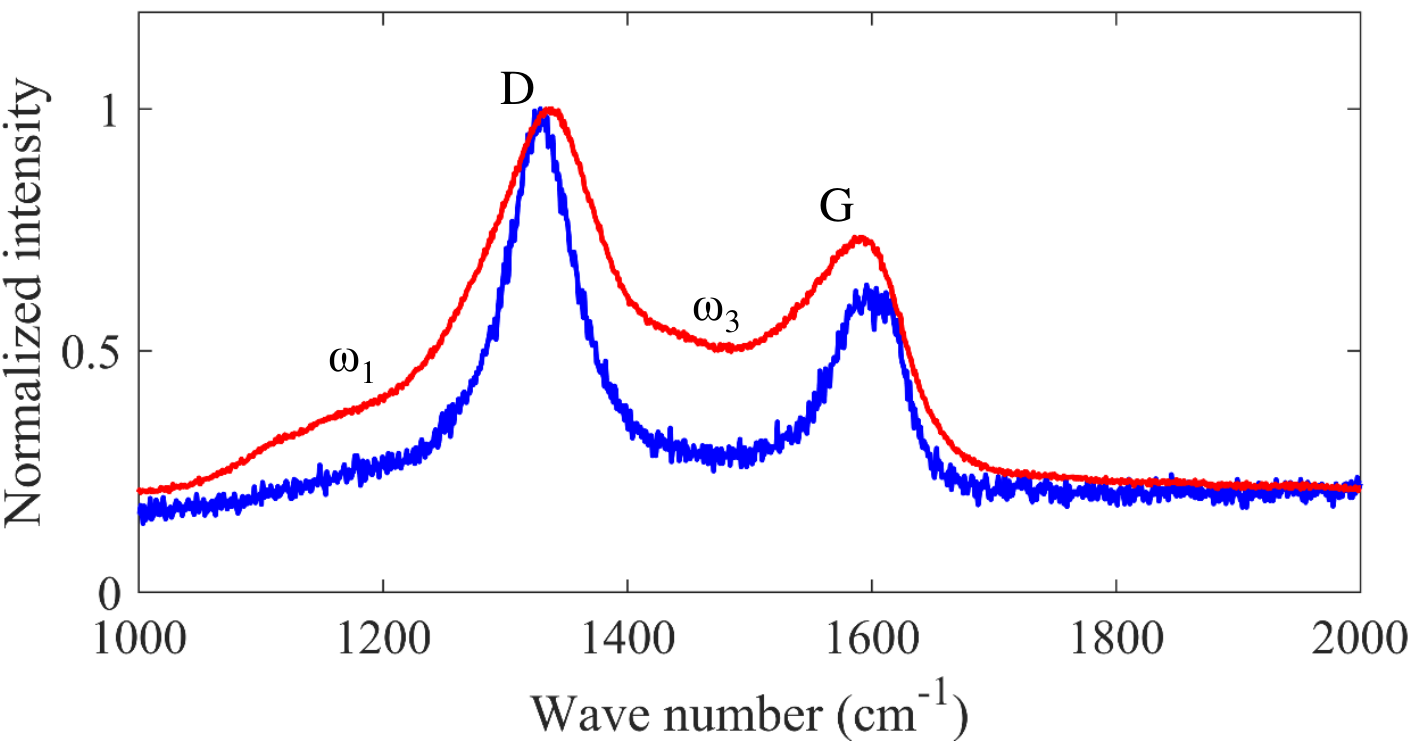}
		\caption{\label{Fig_Raman} Raman spectra taken at different locations across the (N)UNCD surface showing phase variation. Measurements were performed using primary probe wavelength of 633 nm.}
	\end{figure}

	Raman spectroscopy revealed that the graphitic phase increased (as compared to Ref.~\cite{SergeyAPL2014}), but at the same time appeared to non-uniformly distribute across the surface, as illustrated in Fig.~\ref{Fig_Raman}. Some locations (red curve) demonstrated nearly canonical (N)UNCD spectra. Meanwhile, other locations (blue curve) showed stronger graphitization due to higher local temperature which is evidenced by the disappearance of $\omega_1$ and $\omega_3$ bands associated with trans-polyacetylene~\cite{TanviThesis2019}. All characterized locations demonstrate the strong D band  representing the diamond grain $sp^3$ host matrix, and the G band that is associated with the graphitic $sp^2$ grain boundaries. Locations such as that represented by the blue curve in Fig.~\ref{Fig_Raman} are expected to have grain boundaries that are better crystallized compared to locations such as that represented by the red curve.
	
	In spray deposited films comprised of diamond and graphite powders mixed at varied diamond-to-graphite ratios, that are model systems with pre-designed non-uniform graphite-to-diamond distribution, electron emission was found to be non-uniform across the surface~\cite{CuiJVST2002}. Therefore, it could be speculated that such non-uniform graphite-to-diamond ($sp^3$-to-$sp^2$) content distribution is the effect behind non-uniform emission across the nano-diamond surface, as observed in previous dc studies~\cite{KockDiamond2004,ChubenkoACS2017}; both studies showed that surface topography cannot explain enhanced field emission at low turn-on field.
	
	The average work function of the (N)UNCD surface after deposition was measured to be 4.0~eV by a Kelvin probe.

	\section{\label{cha3}Experimental setup}
\begin{figure}[h!tbp]
		\includegraphics[width=8.5cm]{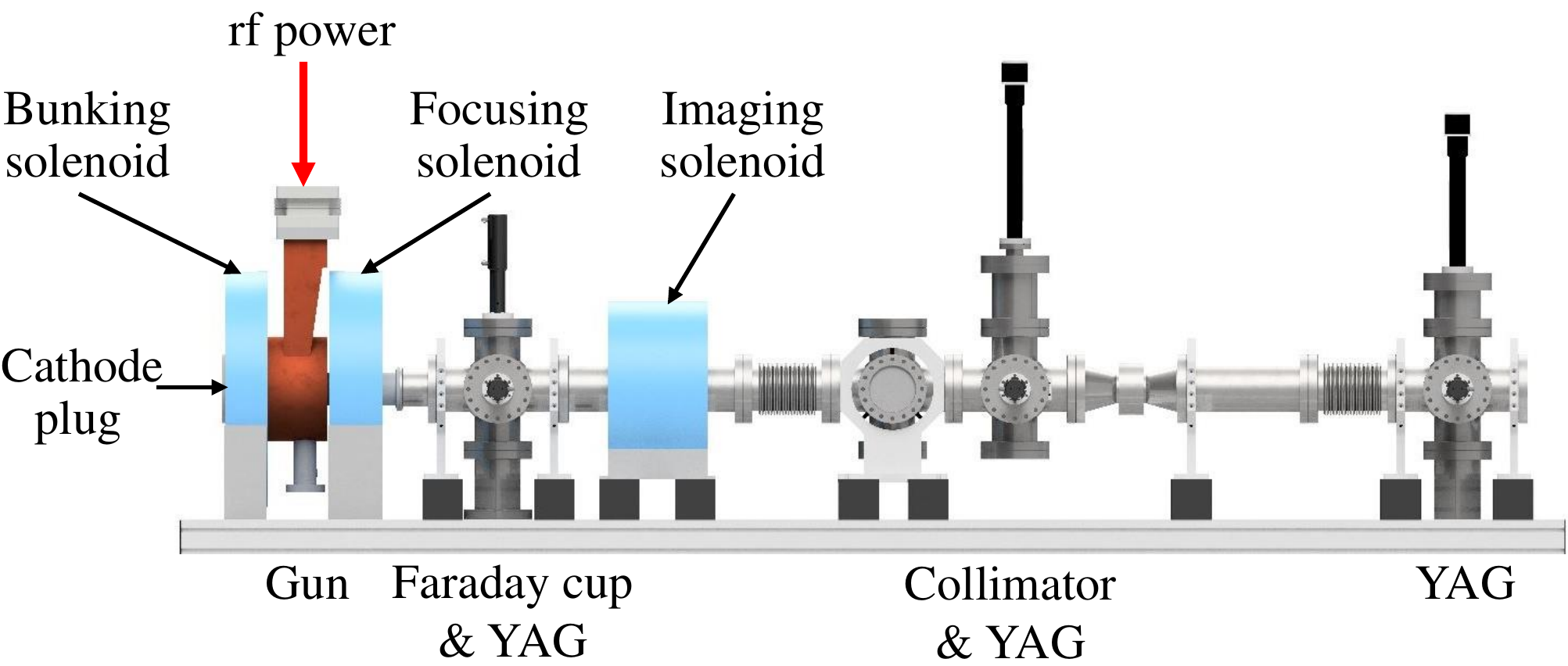}
		\caption{\label{Fig_setup} Layout of the ACT beamline at AWA.}
	\end{figure}

	The experiment was conducted using the Argonne Cathode Test-stand (ACT) beamline at the Argonne Wakefield Accelerator (AWA) facility~\cite{SergeyAPL2014,JiahangPRL2015,JiahangPRL2016,Jiahangthesis2018}, which is illustrated in Fig.~\ref{Fig_setup}. The ACT beamline is equipped with a single-cell normal conducting  photocathode rf gun operated in L-band at 1.3~GHz. The beamline currently runs at a 2~Hz repetition rate with a full width half maximum (FWHM) pulse length of 6~$\mu$s. Cathodes with various shapes and materials can be tested thanks to the detachable cathode design. A frequency tuner also allows the cathodes to be tested at a wide range of longitudinal positions inside the cavity, which results in a drastically different field on the cathode surface for the same input rf power~\cite{JiahangPRL2015}. In this experiment, the cathode surface was set to be flush with the rf gun back wall. At this position, $\sim$227~W input power is required to obtain 1~MV/m cathode electric field. Vacuum in the beamline was maintained below $5 \times 10^{-9}$~Torr.
	
\begin{figure}[h!tbp]
		\includegraphics[width=8cm]{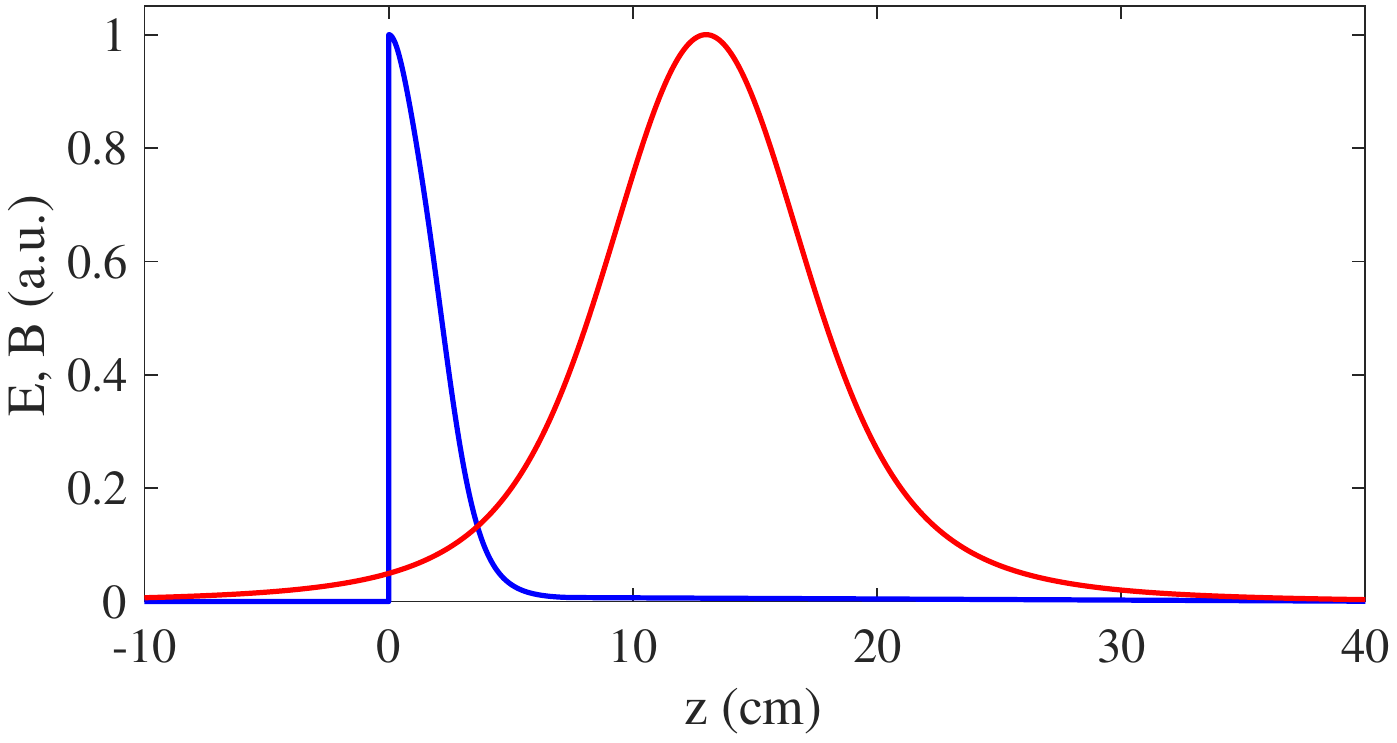}
		\caption{\label{Fig_gun_EB} Longitudinal field profile of the normalized electric field of the rf gun (blue) and the normalized magnetic field of the focusing solenoid (red). The electric field decreases monotonically from the cathode (z=0) to the gun exit (z=7.7~cm) due to the single-cell design.}
	\end{figure}

	Diagnostics involved in the experiment include a direction coupler to monitor the input and reflected rf power, an rf pickup installed at the gun side-wall to detect the field profile, an aluminum block acting as a Faraday cup at the gun exit to collect the field emission current, an imaging system consisting of solenoids and a collimator to obtain the field emitter distribution with high resolution~\cite{JiahangPRL2016,Jiahangthesis2018}, and YAG screens to observe the beam transverse profile along the beamline. A photomultiplier tube (PMT) with a fluorescent screen sensitive to X-rays was placed near the gun for machine interlock purposes in the event of rf breakdown~\cite{JuwenSLAC1997}. The strength of the focusing solenoid (denoted as $B_{f}$) was used to maximize the electron beam capture ratio by the Faraday cup. The longitudinal on-axis field profile of the rf gun and the focusing solenoid are illustrated in Fig.~\ref{Fig_gun_EB}.
	
	\section{\label{cha4}High power test}
	\subsection{Field emission charge collection}
	The capture ratio of the Faraday cup depends on the rf gun field as well as the focusing solenoid strength. Determining the capture ratio under various circumstances is critical to interpret the field emission properties in the measurement. This subsection introduces the simulation and experimental efforts to ensure effective charge collection as well as to determine the capture ratio.
	
	In this subsection, $|E_{c}(t)|$ denotes the macroscopic cathode field amplitude within the 6~$\mu$s rf pulse, $E_{c,\max}$ denotes its maximum value, and $E_{c}(t)=|E_{c}(t)|\cos (\omega t)$ denotes the transient cathode field where $\omega$ is the operation frequency. Since the rf pulse length is much longer than the rf cycle (1/1.3~GHz=769.2~ps), the slow varying of the field amplitude is ignored and $|E_{c}(t)|$ is treated as a constant $|E_{c}|$ in each rf cycle. $\eta _{cycle}$ and $\eta _{pulse}$ denote the capture ratio within one rf cycle and one rf pulse, respectively. 
	
	\subsubsection{Capture ratio within one rf cycle}
\begin{figure}[h!tbp]
		\includegraphics[width=8cm]{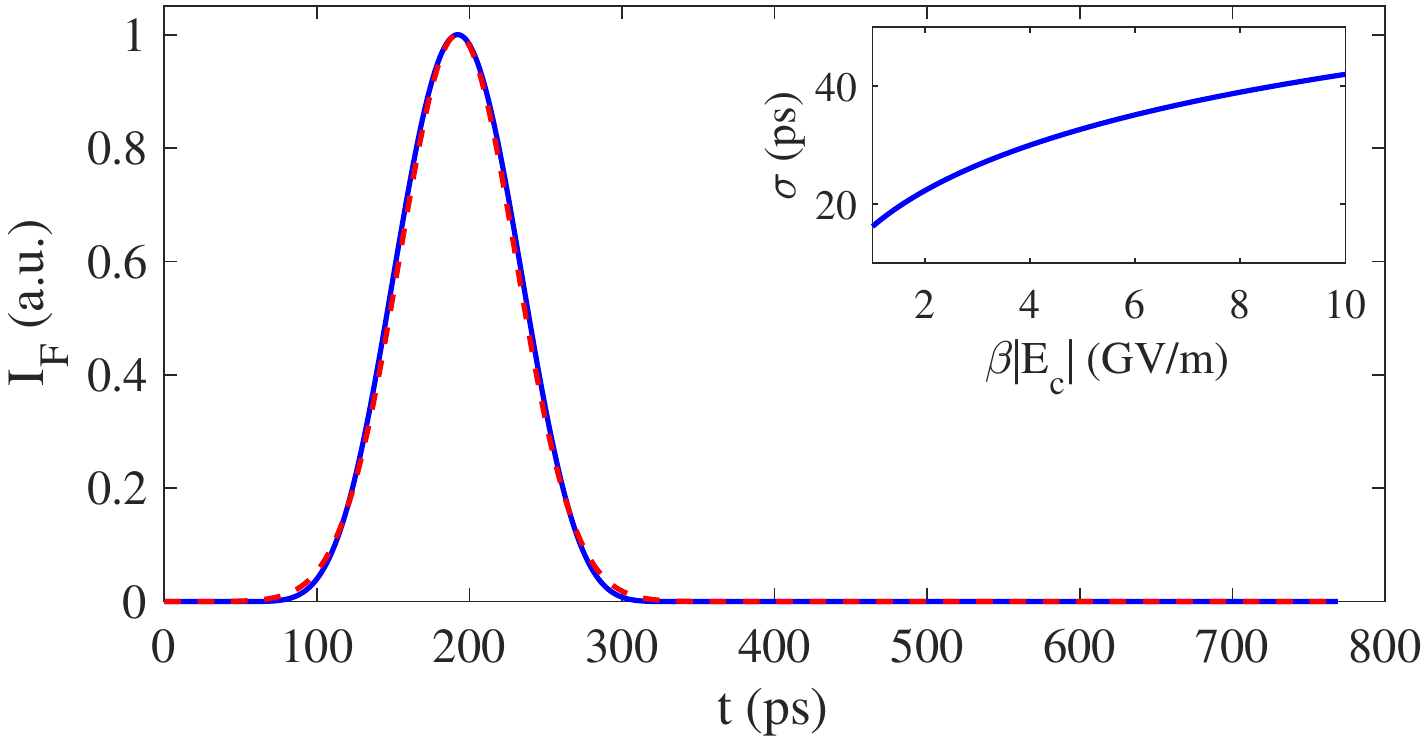}
		\caption{\label{Fig_FE_cycle} Emission profile within one rf cycle based on Eqn.~\ref{Eqn_FN_cos} (blue solid line) and its Gaussian distribution approximation (red dashed line). Inset: The standard deviation of the Gaussian distribution as a function the maximum microscopic cathode field.}
	\end{figure}

	According to the Fowler-Nordheim (F-N) equation, the transient field emission current when the cathode field is positive ($\cos (\omega t)>0$) can be expressed as~\cite{FN1928,JuwenSLAC1997}
	\begin{equation}\label{Eqn_FN_cos}
	\begin{aligned}
	I_{F}(t)=&\frac{1.54\times 10^{-6}\times 10^{4.52\phi ^{-0.5}}A_{e}[\beta E_{c}(t)]^{2}}{\phi}\\
	&\times \exp[-\frac{6.53\times 10^{9}\phi ^{1.5}}{\beta E_{c}(t)}]\\
	\end{aligned}
	\end{equation}
	where $\beta$ is the field enhancement factor, $A_{e}$ is the effective emission area, and $\phi$ is the work function. The emission profile can be approximated by a Gaussian distribution~\cite{RuixuanPRST2015} whose standard deviation $\sigma$ depends on the maximum microscopic cathode field $\beta |E_{c}|$, as illustrated in Fig.~\ref{Fig_FE_cycle}.
	
	As the emission period covers 180$^{\circ}$ in the rf cycle, electrons experience different accelerating fields and focusing strengths depending on their emitting phases. A portion of the field emission electrons can not be captured by the Faraday cup due to the finite gun aperture (40~mm in diameter), the limited Faraday cup size (64~mm tall and 71~mm wide), as well as the beam dynamics inside the rf gun.
	
\begin{figure}[h!tbp]
		\includegraphics[width=8cm]{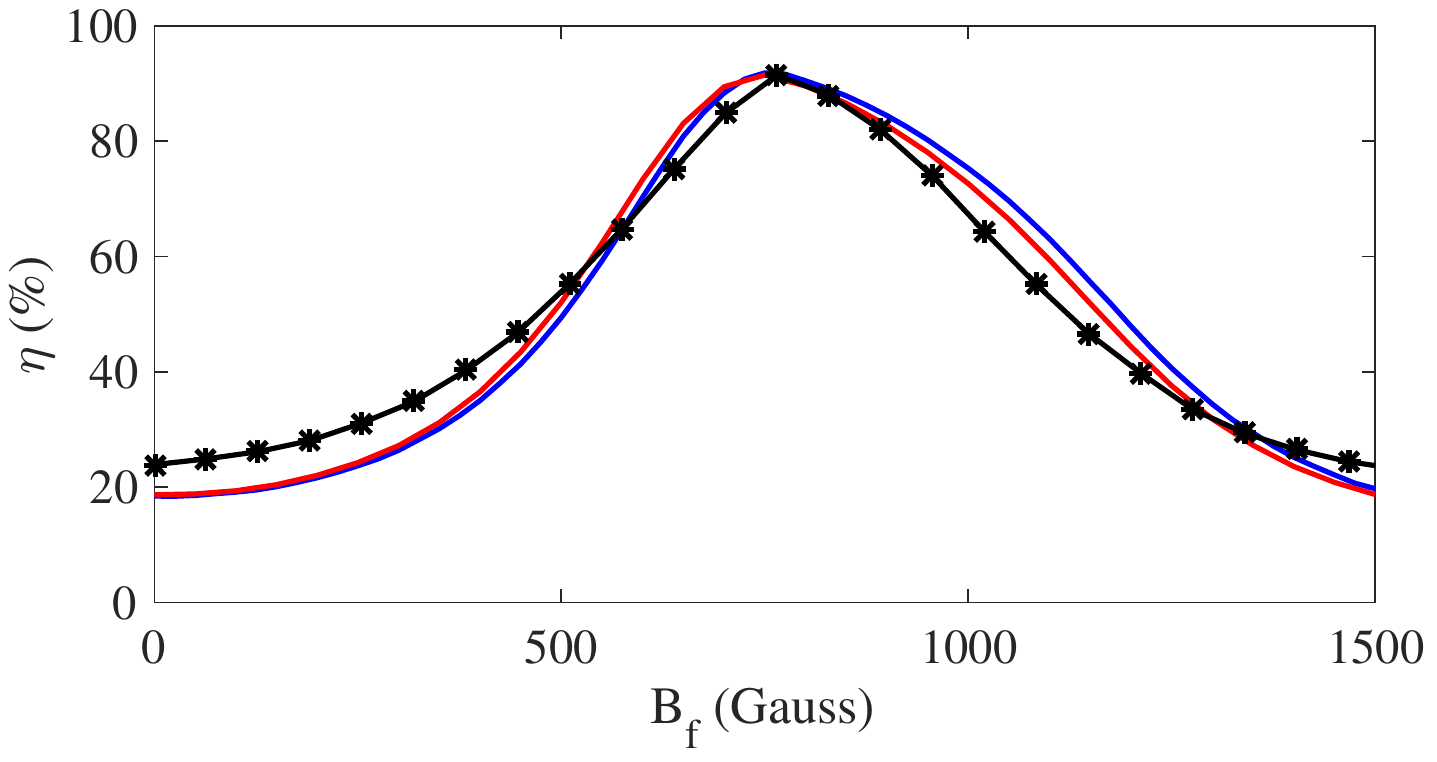}
		\caption{\label{Fig_capture_maxB2} The capture ratio of the Faraday cup with $\beta$ of 179. Blue: $\eta _{cycle}$ with $|E_{c}|$ of 42~MV/m; Red: $\eta _{pulse}$ with $E_{c,\max}$ of 42~MV/m; Black: measured charge at $E_{c,\max}$ of 42~MV/m with the maximum value normalized to the highest capture ratio in simulation.}
	\end{figure}

	Beam dynamics simulations with ASTRA~\cite{ASTRA} have therefore been conducted to examine the capture ratio, in which the electrons were emitted uniformly from the (N)UNCD covered area (18~mm in diameter) on the cathode with an initial thermal kinetic agitation of 0.1~eV. The longitudinal emission profile was set to be Gaussian with $\sigma$ determined by the fitted $\beta |E_{c}|$ (introduced in the following sections).  The blue line in Fig.~\ref{Fig_capture_maxB2} illustrates an example of the simulated capture ratio $\eta_{cycle}$ dependence on focusing solenoid strength B$_{f}$, with $|E_{c}|$ of 42~MV/m, $\beta$ of 179, and $\beta |E_{c}|$ of 7.5~GV/m. It shows that over 90~\% of the emitted electrons can be captured by the Faraday cup when $B_{f}$ is set to 750~Gauss.
	
	\subsubsection{Capture ratio within one rf pulse with fixed $E_{c,\max}$}
\begin{figure}[h!tbp]
		\includegraphics[width=8cm]{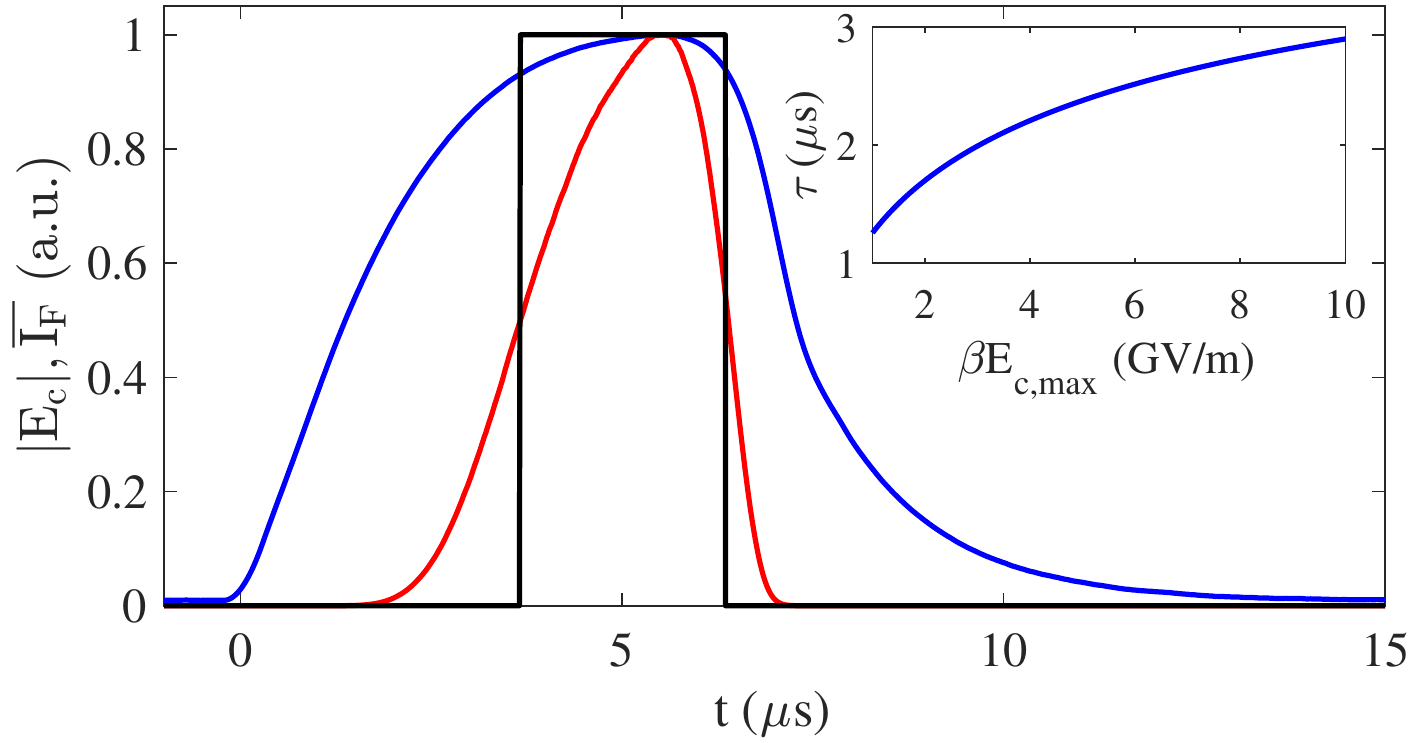}
		\caption{\label{Fig_FE_rf_pulse} Blue: the normalized cathode field amplitude $|E_{c}(t)|$ measured by the rf pickup. Red: the predicted average emission current $\overline{I_{F}(t)}$ by Eqn.~\ref{Eqn_FN_ave}. Black: the square pulse approximation of the emission profile. Inset: The width of the square emission profile as a function of $\beta E_{c,\max}$.}
	\end{figure}

	Within the 6~$\mu$s rf pulse, the macroscopic cathode electric field amplitude varies due to the finite pulse length and the filling time of the standing-wave cavity, as illustrated in Fig.~\ref{Fig_FE_rf_pulse}. Therefore, $\eta _{cycle}$ is also time-dependent and $\eta _{pulse}$ can be defined as 
	\begin{equation}\label{Eqn_eta_pulse}
	\eta _{pulse}=\frac{\int \eta _{cycle}(t) \overline{I_{F}(t)}dt}{\int \overline{I_{F}(t)}dt}
	\end{equation}
	where $\overline{I_{F}(t)}$ denotes the average emission current within one rf cycle that can be expressed as~\cite{JuwenSLAC1997} 
	\begin{equation}\label{Eqn_FN_ave}
	\begin{aligned}
	\overline{I_{F}(t)}=&\frac{5.7\times 10^{-12}\times 10^{4.52\phi ^{-0.5}}A_{e}[\beta |E_{c}(t)|]^{2.5}}{\phi ^{1.75}}\\
	&\times \exp[-\frac{6.53\times 10^{9}\phi ^{1.5}}{\beta |E_{c}(t)|}]\\
	\end{aligned}
	\end{equation}
	
	In ASTRA simulations of the capture ratio within one rf pulse with fixed $E_{c,\max}$, $\eta _{cycle}$ at various field levels were individually studied in the same manner as described in the previous subsection; then $\eta _{pulse}$ was calculated using the predicted emission profile as illustrated in Fig.~\ref{Fig_FE_rf_pulse}. For each $|E_{c}|$, $\beta$ was assumed to be constant and $\sigma$ of the longitudinal emission profile was adjusted based on $\beta |E_{c}|$. The red line in Fig.~\ref{Fig_capture_maxB2} illustrates the $\eta _{pulse}$ dependence on focusing solenoid strength B$_{f}$, with $E_{c,\max}$ of 42~MV/m and $\beta$ of 179, which shows good agreement with the experimental results.
	
	In Fig.~\ref{Fig_capture_maxB2}, the blue and the red lines are very similar to each other, which implies that $\eta _{pulse}$ is dominated by $\eta _{cycle}$ at $E_{c,\max}$. This can be understood since $\overline{I_{F}(t)}$ is highly sensitive to $|E_{c}|$. As illustrated in Fig.~\ref{Fig_FE_rf_pulse}, $\overline{I_{F}(t)}$ drops to 1~\% of its maximum value when $|E_{c}|$ decreases by only 20~\%. The capture ratio study can therefore be simplified by using a square emission profile with an average emission current of $\overline{I_{F,\max}}$ calculated by using $|E_{c}|=E_{c,\max}$ in Eqn.~\ref{Eqn_FN_ave}, as illustrated in Fig.~\ref{Fig_FE_rf_pulse}. The width of the square pulse is set as $\tau =\int \overline{I_{F}(t)}dt/\overline{I_{F,\max}}$ so as to keep the same charge. It depends on $\beta E_{c,\max}$, as calculated by Eqn.~\ref{Eqn_FN_ave} and illustrated in the inset of Fig~\ref{Fig_FE_rf_pulse}.
	
	\subsubsection{Capture ratio within one rf pulse with various $E_{c,\max}$}
	The dependence of $\overline{I_{F,\max}}$ on $E_{c,\max}$ can be plotted in the $1/E_{c,\max}$-$\lg (\overline{I_{F,\max}}/E_{c,\max}^{2.5})$ coordinate (a.k.a. the F-N coordinate). The dependence of $\lg (\overline{I_{F,\max}}/E_{c,\max}^{2.5})$ on $1/E_{c,\max}$ is linear when the field emission is not limited by the space charge effect~\cite{BarbourPR1953,RokhlenkoJAP2010}, from which $\beta$ and $A_{e}$ can be respectively fitted as~\cite{JuwenSLAC1997}
	\begin{equation}\label{Eqn_beta_Ae}
	\left\{
	\begin{aligned}
	\beta&=\frac{-2.84\times 10^{9}\phi ^{1.5}}{s}\\
	A_{e}&=\frac{10^{y_{0}}\phi ^{1.75}}{5.7\times 10^{-12}\times 10^{4.52\phi ^{-0.5}}\beta}\\
	\end{aligned}
	\right.
	\end{equation}
	where $s$ and $y_{0}$ are the slope and the y-axis intercept of the linear dependence.
	
	Experimental studies of emission properties usually record various electric field levels and the corresponding field emission current in order to fit $\beta$ and $A_{e}$. The capture ratio under these field levels may not be the same especially in rf structures with complicated geometries, which could lead to inaccurate results. This effect, however, was seldom considered in previous research.
	
\begin{figure}[h!tbp]
		\includegraphics[width=8cm]{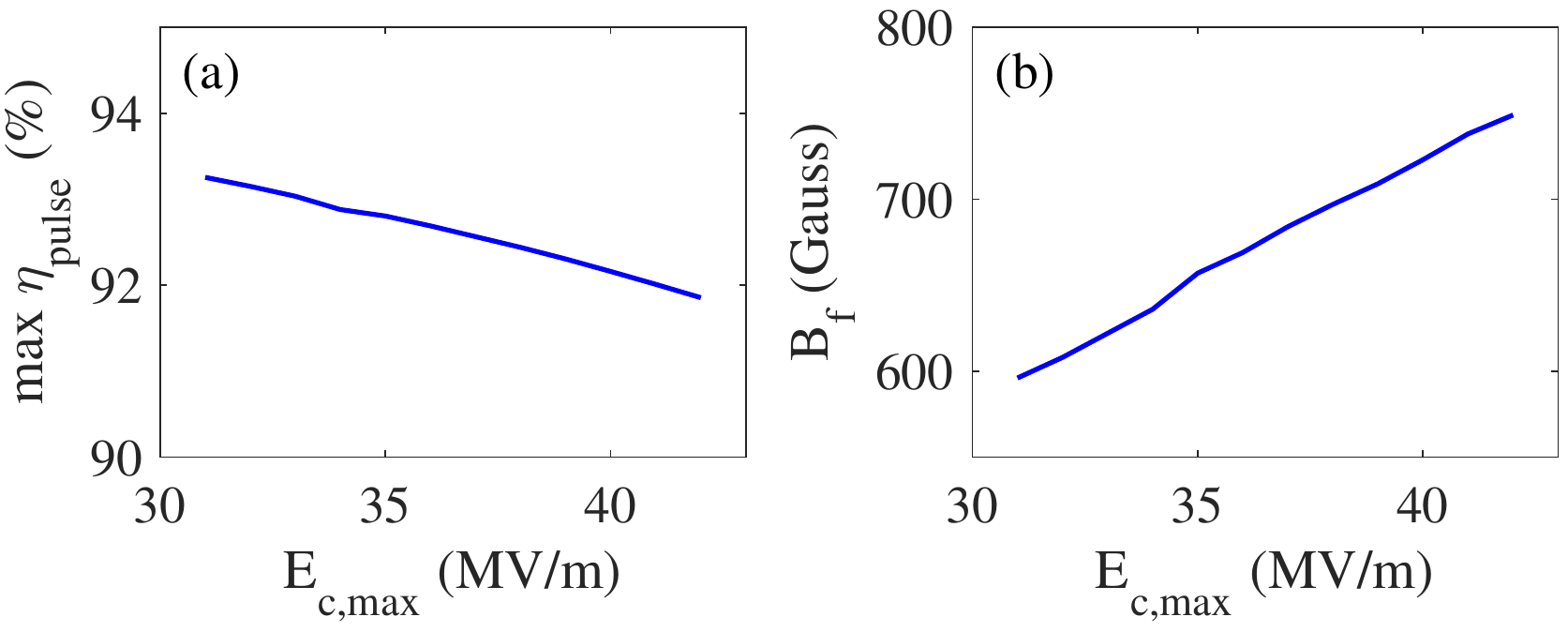}
		\caption{\label{Fig_capture_maxE} The maximum $\eta _{pulse}$ (a) and the corresponding $B_{f}$ (b) as a function of $E_{c,\max}$. In this simulation, $\beta$ is fixed at 179.}
	\end{figure}

	This effect has been carefully simulated in the presented study. As described in the previous subsection, the emission profile in this simulation step was also approximated by a square pulse with its width determined by $\beta E_{c,\max}$. The value of $\beta$ paired with each $E_{c,\max}$ was kept constant with the assumption that the surface condition doesn't change during the measurement. This assumption is valid since measurements usually took less than 300~rf pulses (2.5~minutes) without any rf breakdown, during which the surface condition evolution was negligible compared to the entire experiment period of tens of hours with thousands of rf breakdowns. The variation of $E_{c,\max}$ in simulation was set to be 20~\%, close to the one used in experiment which was limited by the minimal detectable charge of the Faraday cup. The simulation results, as illustrated in Fig.~\ref{Fig_capture_maxE}, suggest that similar capture ratio (with a maximum difference of less than 2~\%) can be achieved by adjusting $B_{f}$, which supports the accuracy of the fitted results in the following sections.
	
	\subsection{Field emitter distribution}
	The field emitter distribution on the (N)UNCD cathode was studied with low resolution by taking regular YAG images at the gun exit and with high resolution using the \textit{in-situ} field emission imaging system at the downstream end of the beamline.
	
	\subsubsection{Low resolution observation at the gun exit}
	Due to the detachable cathode design, field emission electrons may come from the edge of the cathode/insertion hole rather than the cathode itself. The two sources can not be distinguished by the Faraday cup. Therefore, a YAG screen (denoted as YAG$_{1}$) located at the same location as the Faraday cup was used to evaluate the emission current from these sources.
	
\begin{figure}[h!tbp]
		\includegraphics[width=8cm]{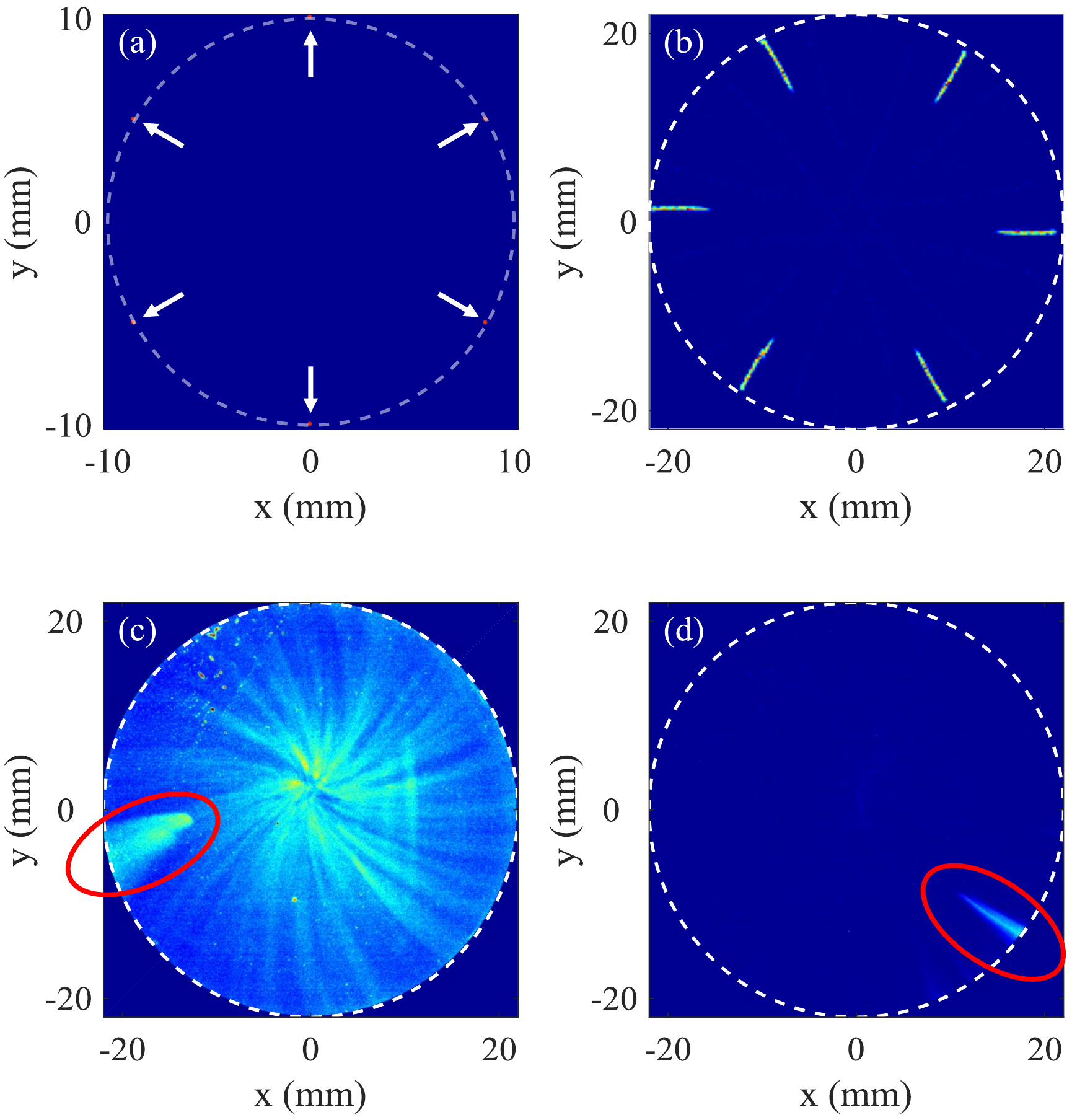}
		\caption{\label{Fig_YAG1} (a) The initial edge emitters distribution used in the ASTRA simulation. The white dashed circle represents the $\Phi$20~mm cathode edge and the white arrows point to the six emitters. (b) Simulated transverse distribution of field emission electrons from the six emitters on YAG$_{1}$. (c) Experimental observation with the (N)UNCD cathode on YAG$_{1}$. (d) Experimental observation with a molybdenum cathode on YAG$_{1}$. In (b-d), the images were simulated or taken with the same $E_{c,\max}$ and $B_{f}$ settings. The white dashed circle represents the $\Phi$44~mm YAG boundary. The red circles mark electrons from edge emitters.}
	\end{figure}

	ASTRA simulations have been conducted to predict the transverse electron distribution at the YAG$_{1}$ position. Six $\Phi$0.2~mm field emitters were placed on the cathode edge in simulation, as illustrated in Fig.~\ref{Fig_YAG1}(a). Electrons emit uniformly within each emitter with an initial kinetic energy of 0.1~eV. They emit during 180$^{\circ}$ of the rf phase when the electric field is positive, and the longitudinal profile during one rf cycle has a Gaussian distribution as described in previous sections. The simulated YAG$_{1}$ image is illustrated in Fig.~\ref{Fig_YAG1}(b). The line-shaped pattern is caused by the wide energy spread of the beam from the broad emission phase, which is common for field emission in rf guns~\cite{HanPAC2005,MorettiPRST2005,DowellPAC2007,XiangPRST2014,JiahangPRL2016}. 
	
	The simulation results are compared with the experimental observation under the same $E_{c,\max}$ and $B_{f}$ to identify the source of emission current. With the (N)UNCD cathode, the line-shaped pattern marked by the red circle in Fig.~\ref{Fig_YAG1}(c) has a similar shape and location as that caused by an edge emitter in the ASTRA simulation, which indicates that there was one emitter at the  edge of the cathode/insertion hole. The rest of the observed pattern on YAG$_{1}$ is caused by emitters on the (N)UNCD material, whose total brightness is much higher than that of the edge emitter. Therefore, the collected charge should be dominated by (N)UNCD emitters rather than the edge emitter. In comparison, the experimental result using a molybdenum cathode without (N)UNCD material clearly shows that the emission was mainly from edge emitters, as illustrated in Fig.~\ref{Fig_YAG1}(d).
	
	\subsubsection{High resolution observation at downstream of the beamline}
\begin{figure}[h!tbp]
		\includegraphics[width=8cm]{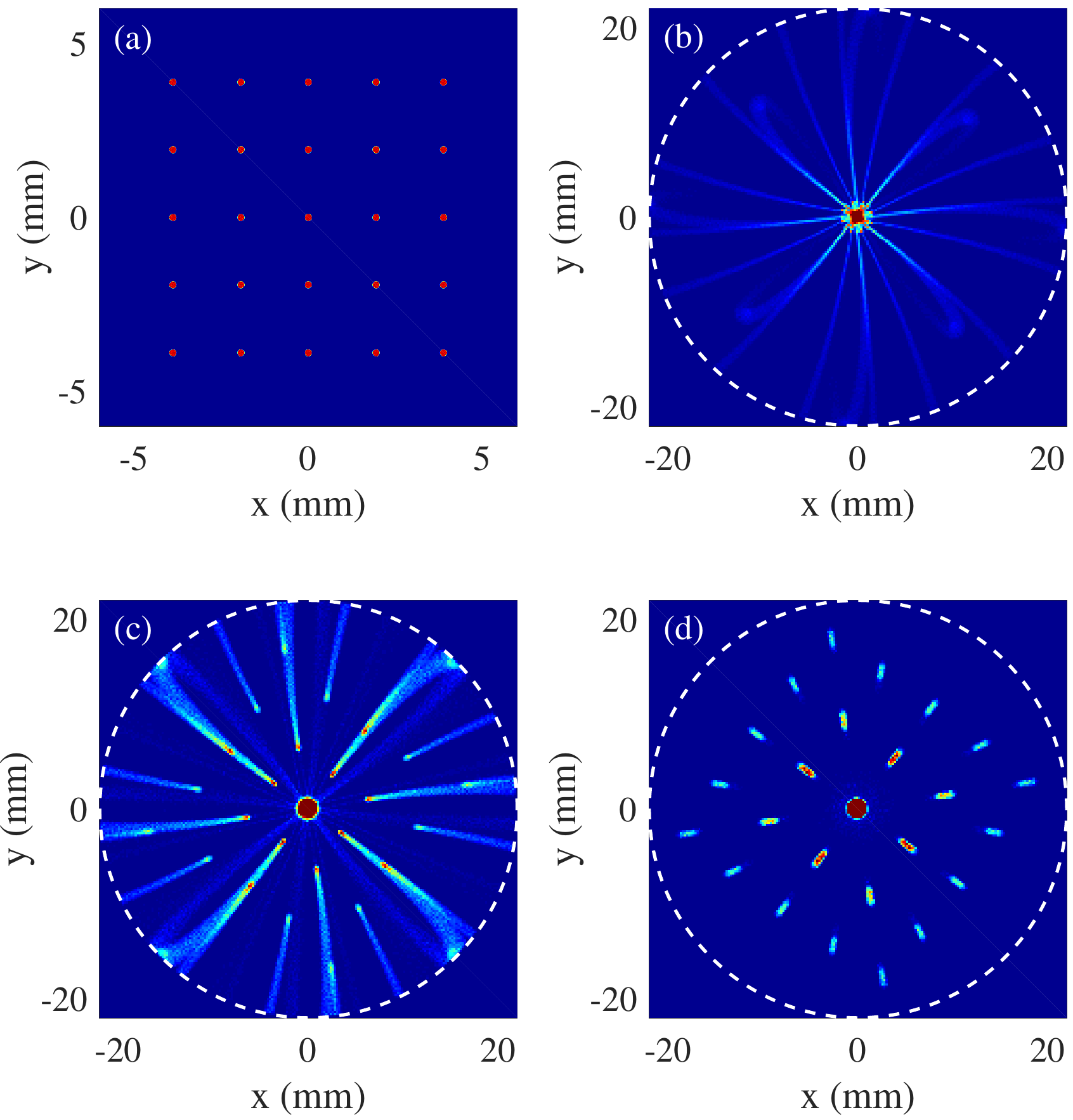}
		\caption{\label{Fig_imaging_simulation} ASTRA simulation results of the \textit{in-situ} field emission imaging system. (a) Field emitter distribution on the cathode, which consists of 25 $\Phi$0.2~mm emitters separated by 1.77~mm. (b) Electron transverse distribution at the collimator position. (c) Electron transverse distribution on YAG$_{3}$ without applying the collimator. (d) Electron transverse distribution on YAG$_{3}$ with a $\Phi$1~mm collimator. In (b-d), the white dashed circles represent the $\Phi$44~mm YAG boundary.} 
	\end{figure}

	The high resolution field emission imaging system~\cite{JiahangPRL2016,Jiahangthesis2018} was applied to improve the resolution of the field emitter distribution on the cathode. In this system, a solenoid (the imaging solenoid in Fig.~\ref{Fig_setup}, with strength denoted as $B_{i}$) is used to focus the beam. The focal length depends on the beam energy, which is determined by the emitting phase. Electrons with certain energies (emitting phases) can be selected by placing a collimator with a small aperture after the solenoid. The transverse electron distribution on the last YAG screen of the beamline (denoted as YAG$_{3}$) can be used to reproduce the field emitter distribution on the cathode with certain magnification and rotation. The ASTRA simulation study illustrated in Fig.~\ref{Fig_imaging_simulation} demonstrates the working principle of the imaging system.
	
	The resolution of the system determines the capability of distinguishing emitters on the cathode. Due to the axial symmetry of the imaging system, the resolution can be defined in the radial and angular directions (denoted as $R_{\rho}$ and $R_{\varphi}$). Assuming electrons of a zero-size emitter, after collimation they follow Gaussian distributions in both the radial and angular directions with standard deviations of $\delta _{\rho}$ and $\delta _{\varphi}$. The resolutions of the imaging system can be expressed as~\cite{JiahangPRL2016,Jiahangthesis2018}
	\begin{equation}
	\label{Eqn.resolution}
	\left\{
	\begin{aligned}
	R_{\rho}&=2.35 \times \dfrac{\delta _{\rho}}{\overline{mag}}\\
	R_{\varphi}&=2.35 \times \delta _{\varphi}\rho _{0}\\
	\end{aligned}
	\right.
	\end{equation}
	where $\overline{mag}$ is the average magnification and $\rho _{0}$ is the emitter radial position on the cathode. 
	
	Under the experimental conditions ($E_{c,\max}$ of 36~MV/m, $B_{f}$ of 750~Gauss, $B_{i}$ of 250~Gauss, and the aperture diameter of 1~mm), $R_{\rho}$, $R_{\varphi}$, and $\overline{mag}$ are simulated to be $\sim$300~$\mu$m, $\sim$10~$\mu$m, and 3.7, respectively. It should be noted that the resolution is not adequate to resolve the fine structure of (N)UNCD emitters whose sizes were determined to be on the order of $\mu$m and below in previous dc studies~\cite{CuiJVST2002,KockDiamond2004,ChubenkoACS2017}. The achievable resolution in the experiment could be worse than in the simulation due to the camera resolution, the low signal-to-noise ratio images, the rf amplitude jitter, the variation of $E_{c}$ within the rf pulse, etc. As shown in the experiment, however, the \textit{in-situ} imaging system is valuable for distinguishing isolated emitters on the cathode.
	
\begin{figure}[h!tbp]
		\includegraphics[width=8cm]{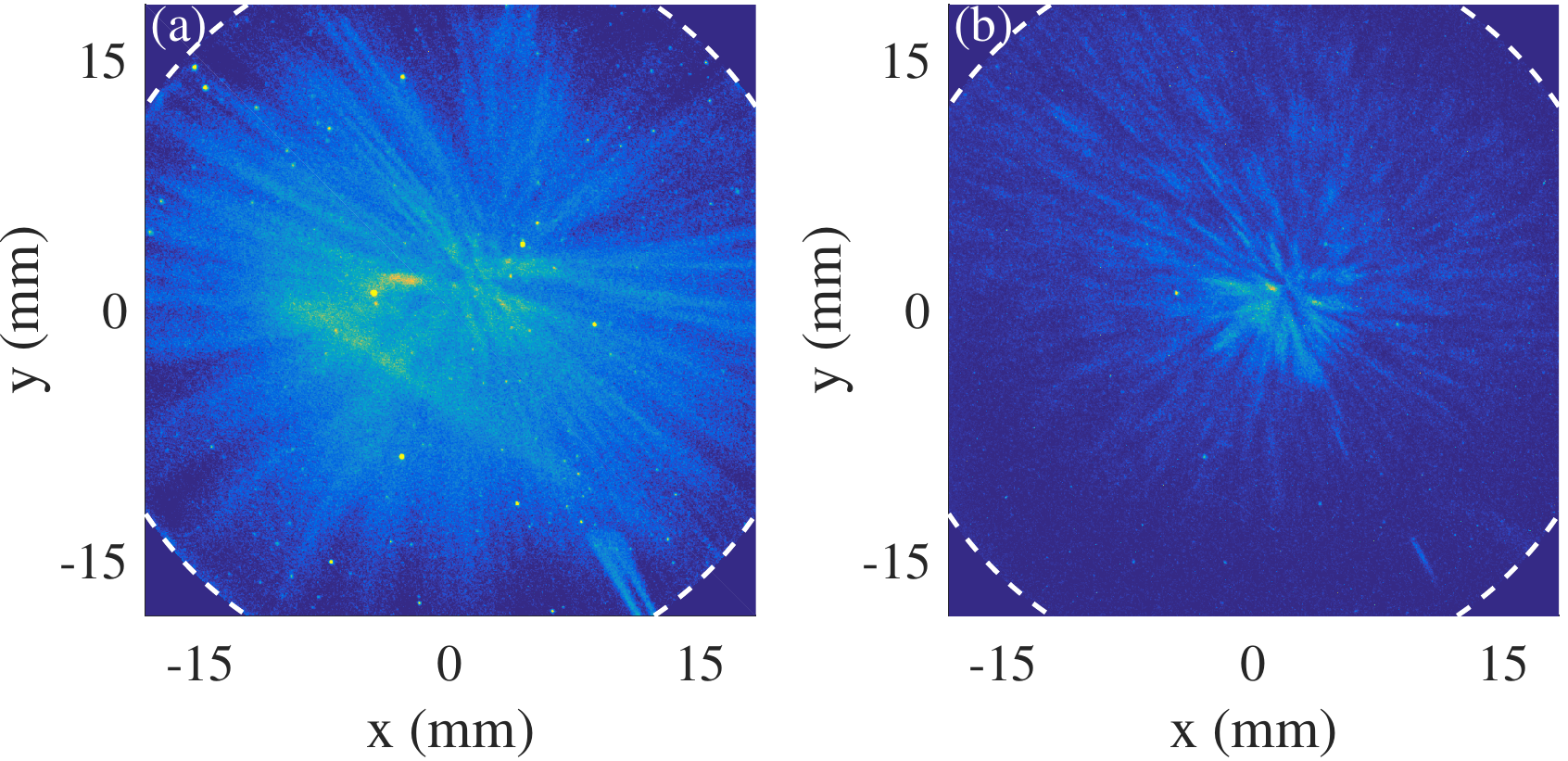}
		\caption{\label{Fig_YAG3} Field emission images obtained on YAG$_{3}$ without collimator (a) and with a $\Phi$1~mm aperture (b). The white dashed circle represent the $\Phi$44~mm YAG boundary.} 
	\end{figure}

	Field emission images on YAG$_{3}$ were taken with the high resolution \textit{in-situ} field emission imaging system when cathode field was conditioned to 36~MV/m, as illustrated in Fig.~\ref{Fig_YAG3}. Each short line-shaped pattern after applying the collimator is caused by a single emitter as suggested in the ASTRA simulation. The observation clearly shows that the field emission current was not uniform from the entire (N)UNCD cathode, but had contributions from a large number of isolated emitters. Given the imaging magnification of 3.7 and YAG diameter of 44~mm, the cathode range that can be observed on YAG$_{3}$ is $\sim \Phi$11.9~mm. Over 100 emitters can be found within this range by counting the short lines, which corresponds to  an emitter density above 100/cm$^{2}$. The isolated emitter behavior and the emitter density agrees with previous observations in dc setups~\cite{CuiJVST2002,KockDiamond2004,ChubenkoACS2017}.
	
	\subsection{rf conditioning}
	The field emission properties of  (N)UNCD were studied during rf conditioning by gradually increasing the gun field, $E_{c,\max}$, from $\sim$8~MV/m to $\sim$42~MV/m. The increment of $E_{c,\max}$ was $\sim$0.5~MV/m in the conditioning process. The breakdown rate could be as high as $1 \times 10^{-1}$~/pulse immediately after the increment. $E_{c,\max}$ was not raised until the breakdown rate decreased to $1 \times 10^{-3}$~/pulse. When continuous breakdown occurred, the field was reduced to a much lower level until breakdown stopped and then the field was pushed back to the breakdown threshold. The entire rf conditioning and measurements lasted for 40~hours.
	
\begin{figure}[h!tbp]
		\includegraphics[width=8cm]{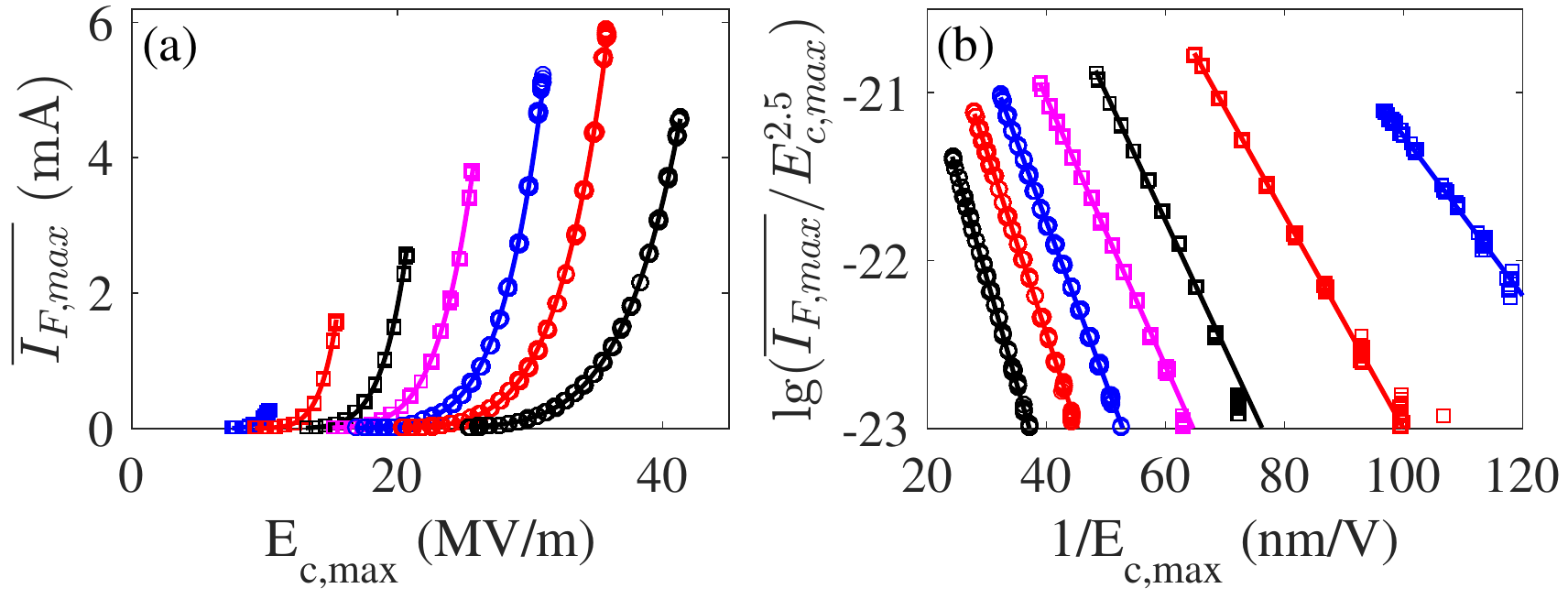}
		\caption{\label{Fig_conditioning} The dependence of $\overline{I_{F,\max}}$ on $E_{c,\max}$ during rf conditioning, plotted in the $E_{c,\max}-\overline{I_{F,\max}}$ coordinate (a) and the F-N coordinate (b). The dots denote the experimental data and the lines represent the linear fitting results from the F-N coordinate. The seven curves, from left to right in (a) and from right to left in (b), were taken when $E_{c,\max}$ reached 10~MV/m, 15~MV/m, 21~MV/m, 26~MV/m, 31~MV/m, 36~MV/m, and 42~MV/m, respectively.} 
	\end{figure}

	When conditioned to certain $E_{c,\max}$ levels, e.g. 20~MV/m, 40~MV/m, etc., $E_{c,\max}$ was maintained until the rf breakdown rate dropped below $5 \times 10^{-4}$~/pulse. Then $E_{c,\max}$ was gradually decreased and $B_{f}$ was adjusted at each $E_{c,\max}$ to maximize the captured current. According to the aforementioned simulation analysis, the maximum capture ratio remains nearly unchanged with $E_{c,\max}$ which ensure the accuracy when fitting $\beta$ and $A_{e}$. The dependence of $\overline{I_{F,\max}}$ on $E_{c,\max}$ is illustrated in Fig.~\ref{Fig_conditioning}. The good linearity in the F-N coordinate suggests that the emission was not space-charge limited~\cite{BarbourPR1953,RokhlenkoJAP2010}. It can be seen that $\overline{I_{F,\max}}$ at a fixed field level kept decreasing during rf conditioning. Meanwhile, the maximum achievable $\overline{I_{F,\max}}$ first rose during the conditioning process, reached $\sim$6~mA at $E_{c,\max}$=36~MV/m, then dropped to $\sim$5~mA at $E_{c,\max}$=42~MV/m.
	
\begin{figure}[h!tbp]
		\includegraphics[width=8cm]{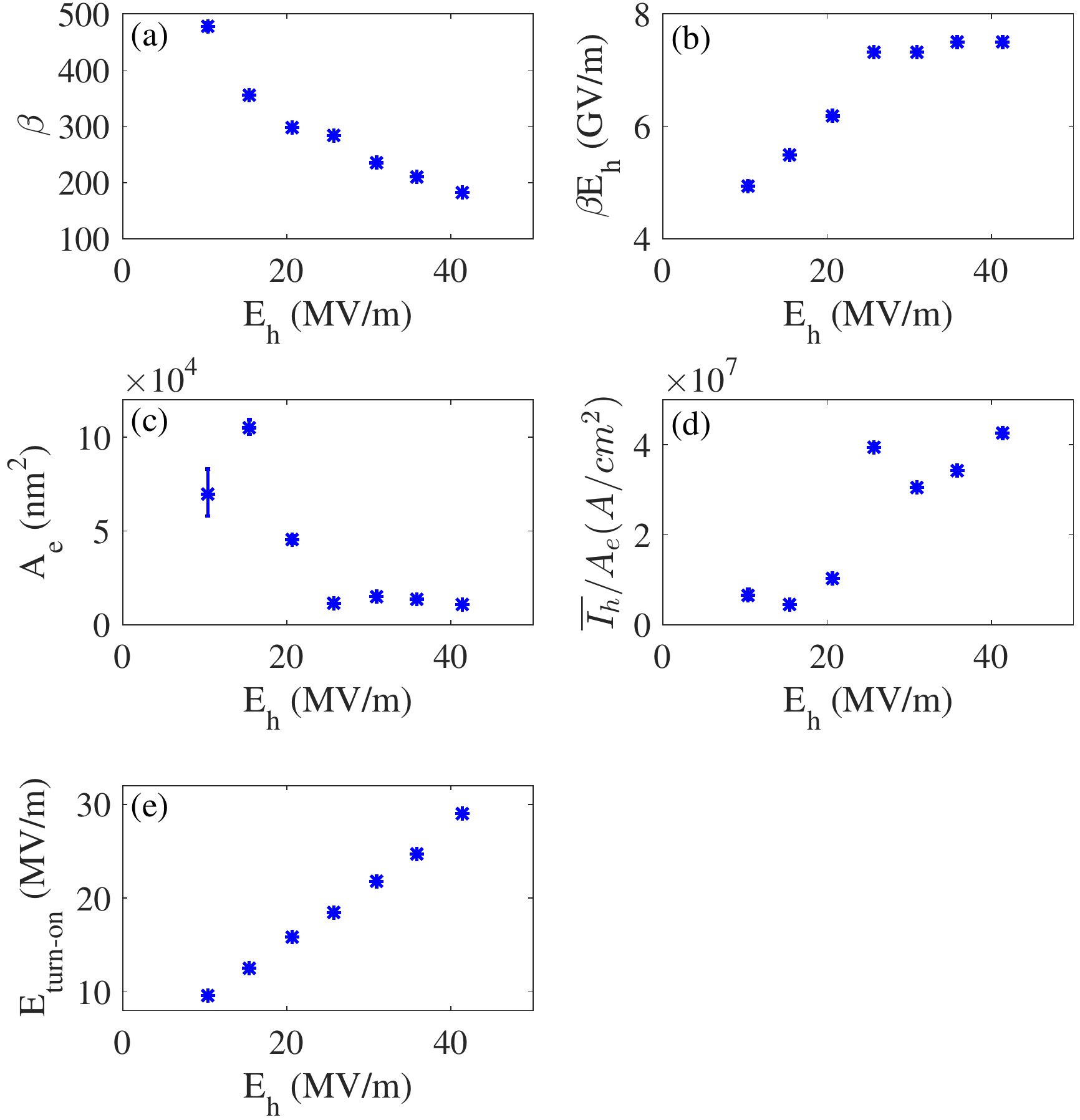}
		\caption{\label{Fig_conditioning_properties} Evolution of field emission properties during rf conditioning. (a) The field enhancement factor; (b) The maximum microscopic electric field; (c) The effective emission area; (d) The current density; (e) The turn-on field.} 
	\end{figure}

	$\beta$ and $A_{e}$ were fitted from the experimental results based on Eqn.~\ref{Eqn_beta_Ae} by assuming $\phi$=4.0~eV. The microscopic field level, the current density, and the turn-on field (defined as the macroscopic field when $\overline{I_{F,\max}}$ reaches 0.1~mA and denoted as $E_{turn-on}$) were derived accordingly, as illustrated in Fig.~\ref{Fig_conditioning_properties}. In the figure, $E_{h}$ and $\overline{I_{h}}$ denote the highest achieved $E_{c,\max}$ and the corresponding $\overline{I_{F,\max}}$, respectively. $\beta$ kept decreasing while the turn-on field kept increasing during the rf conditioning while the maximum microscopic field level $\beta E_{h}$ first increased and then reached a stable level of $\sim$7.5~GV/m. This value agrees with the one in a previous study where $E_{h}$ reached $\sim$70~MV/m~\cite{SergeyAPL2014}. $A_{e}$ decreased by one order of magnitude to $1 \times 10^{4}$~nm$^{2}$ and the current density reached $\sim 4 \times 10^{7}$~A/cm$^{2}$. The behavior of $\beta$ and the turn-on field is analyzed in the discussion section. 
	
	\subsection{Longevity}
\begin{figure}[h!tbp]
		\includegraphics[width=8cm]{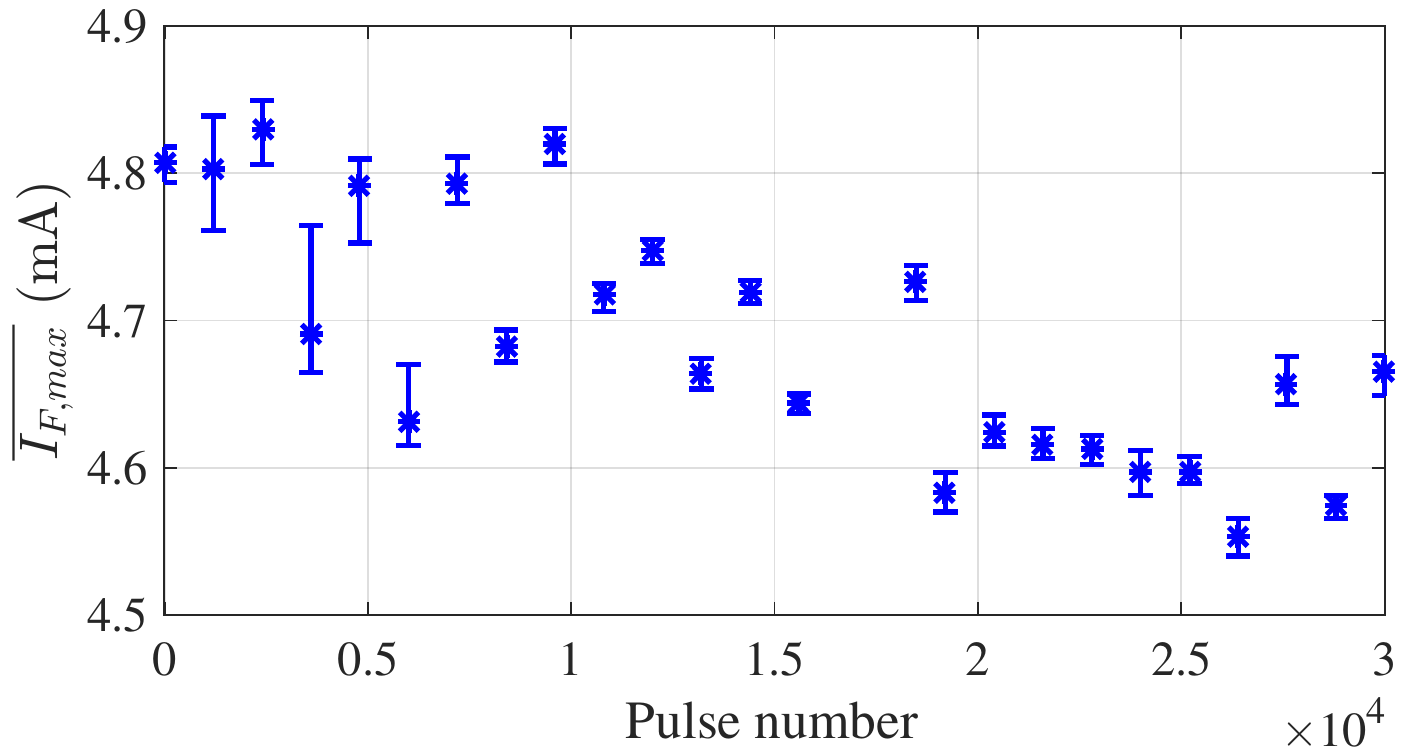}
		\caption{\label{Fig_lifetime} The current evolution during the 4-hour longevity measurement.} 
	\end{figure}

	The longevity of the (N)UNCD cathode was tested when $E_{c,\max}$ reached 42~MV/m. In the 4-hour measurement, $\sim 3 \times 10^{4}$ rf pulses or $\sim 1 \times 10^{8}$ rf cycles (calculated from the 2.5~$\mu$s square emission pulse and 1.3~GHz frequency) were accumulated and the current dropped by only $\sim$4~\%, as illustrated in Fig.~\ref{Fig_lifetime}. Only one breakdown occurred during the measurement. The good longevity and low rf breakdown rate ($3 \times 10^{-5}$~/pulse) demonstrate the promising potential of (N)UNCD material in rf injector applications.
	
\begin{figure}[h!tbp]
		\includegraphics[width=8cm]{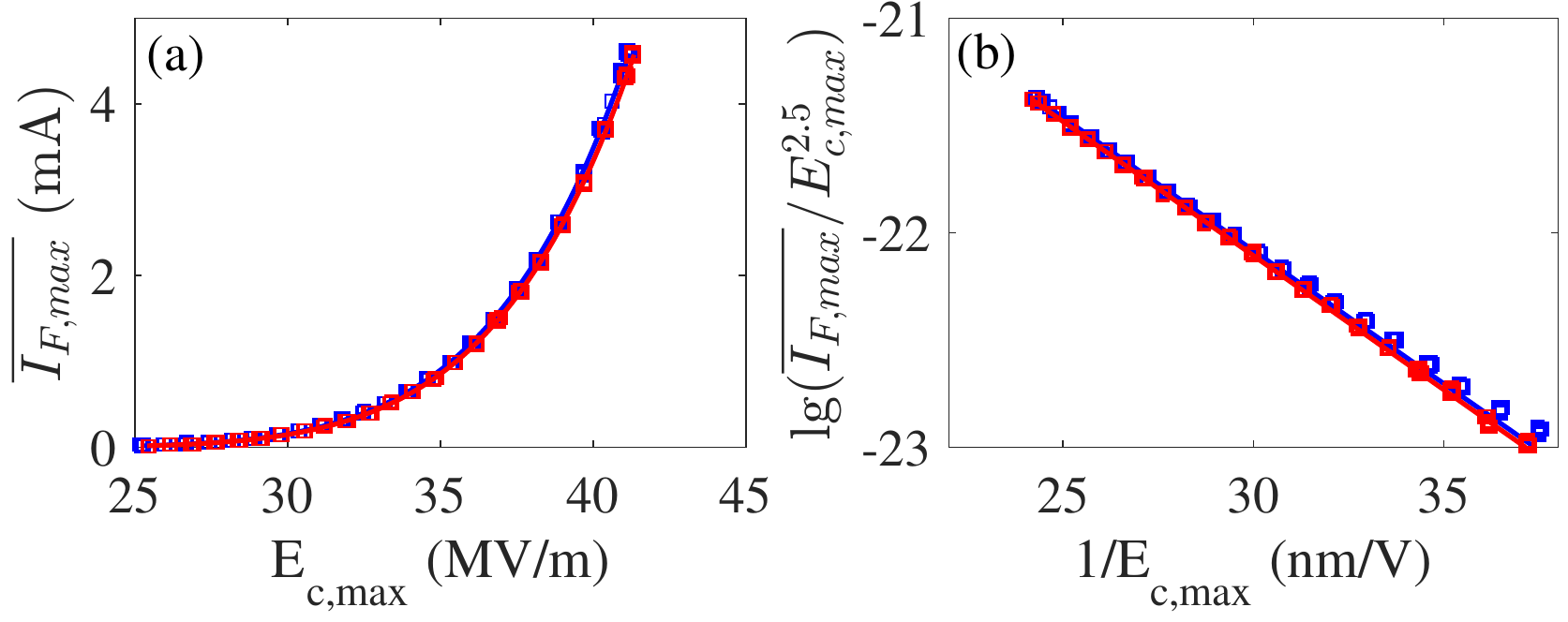}
		\caption{\label{Fig_lifetime_properties} The dependence of $\overline{I_{F,\max}}$ on $E_{c,\max}$ before (blue) and after (red) the longevity measurement, plotted in the $E_{c,\max}-\overline{I_{F,\max}}$ coordinate (a) and the F-N coordinate (b). The dots denote the experimental data and the lines represent the linear fitting results from the F-N coordinate.} 
	\end{figure}

	The field emission properties was measured before and after the 4-hour measurement and the difference is negligible, as illustrated in Fig.~\ref{Fig_lifetime_properties}. Together with the significant variation during rf conditioning, it suggests that the evolution of the field emission properties was mainly caused by rf breakdowns rather than accumulated rf pulses.
	
	\section{\label{cha5}Discussion}	
	In this section, the band diagram of the (N)UNCD material is discussed to interpret the experimental phenomena observed in this study: non-uniform graphite-to-diamond content distribution revealed by the Raman spectra measurement, non-uniform field emitter distribution revealed by the \textit{in situ} imaging system, 3-fold decrease of $\beta$ and 3-fold increase of $E_{turn-on}$ revealed during rf conditioning, and the different surface variation behavior between the conditioning process (remarkable changes of emission properties) and the longevity measurement (nearly unchanged emission properties).  
	
\begin{figure}[h!tbp]
		\includegraphics[width=8cm]{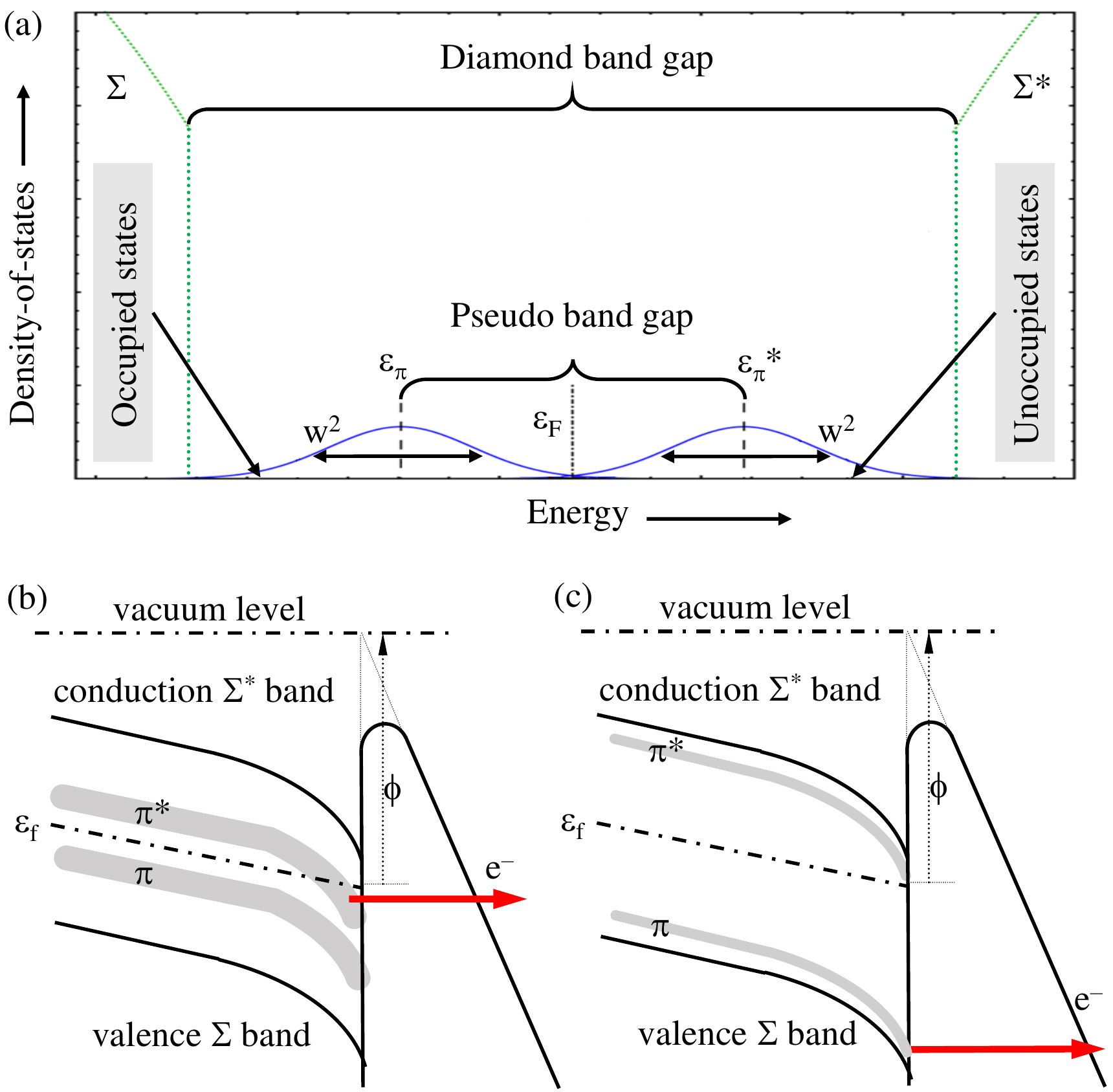}
		\caption{\label{Fig.DOS} (a) Density-of-states diagram of (N)UNCD. Surface barrier under the same external macroscopic electric field for $sp^2$--rich sites (b) and $sp^2$--poor sites (c). The complete self-consistently solved theory based on Stratton-Baskin-Lvov-Fursey formalism can be found in Ref.~\cite{ChubenkoJAP2019}.} 
	\end{figure}

	The basic band diagram of (N)UNCD, $viz.$ the density-of-states (DOS) as a function of energy is illustrated in Fig.~\ref{Fig.DOS}(a). It is modeled as a combination of $sp^2$ (i.e. graphitic grain boundary) induced $\pi$ bands inserted into the fundamental band gap of diamond (i.e. host matrix of diamond grains which are $sp^3$-bonded carbon). The states in the $\pi-\pi^{*}$ bands can be approximated by Gaussian functions centered at $\varepsilon_{\pi}$ and $\varepsilon_{\pi^{*}}$ with variance $w^2$. It is assumed that $\pi-\pi^*$ and $\Sigma-\Sigma^{*}$ bands are mirrored with respect to the Fermi level that is pinned inside the pseudo band gap formed by the $\pi-\pi^{*}$ states. In synthesis, addition of nitrogen to UNCD always adds more graphitic phase~\cite{NesladekPRB1996,ZapolPRB2001}: more states in the $\pi-\pi^{*}$ bands with increased intensity and width; and smaller distance between $\varepsilon_{\pi}$ and $\varepsilon_{\pi^{*}}$. The general trend to expect then is that the overlap between $\pi-\pi^{*}$ bands increases. Consequently, the pseudo band gap becomes smaller (optically darker films~\cite{NesladekPRB1996}) which yields highly conductive semimetallic (N)UNCD.
	
	(N)UNCD is compositionally non-uniform (as illustrated in Fig.~\ref{Fig_Raman}) and two extreme cases are considered for simplicity: $sp^2$--rich (N)UNCD sites in Fig.~\ref{Fig.DOS}(b) and $sp^2$--poor (N)UNCD sites in Fig.~\ref{Fig.DOS}(c). The partial penetration of the electric field into the material leads to band bending near the surface. In $sp^2$--rich (N)UNCD, the $\pi^{*}$ band can be therefore lower than the Fermi level in the near-surface regime. Given the strong overlap (enhanced charge exchange) between $\pi-\pi^{*}$ bands, a large number of electrons is expected to collect near the surface ($metalizing$ it), and to tunnel through the surface barrier with a larger probability as they have higher energy positions inside the material as compared to the states normally occupied without external field perturbation. It results in high charge carrier concentration and metallic-like behavior which further leads to a high field enhancement factor and low turn-on field~\cite{ChubenkoJAP2019}. In $sp^2$--poor (N)UNCD, on the other hand, electrons could only come from the $\pi$ (or $\Sigma$) band with lower probability of tunneling through the thicker surface barrier, resulting in a low field enhancement factor and high turn-on field~\cite{ChubenkoJAP2019}. Under the same macroscopic field, the current density of $sp^2$--rich (N)UNCD should be higher than that of $sp^2$--poor (N)UNCD. This speculation is consistent with the non-uniform emitter distribution revealed by the \textit{in situ} field emission images.
	
	Therefore, the mechanism of (N)UNCD field emission in rf conditioning can be interpreted as follows. 1) Starting at low surface field, emission current is dominated by $sp^2$--rich (N)UNCD sites which causes high $\beta$ and low $E_{turn-on}$ in measurement. 2) When the surface field is increased in the conditioning progress, high localized emission current leads to thermally driven degradation and extinction of originally strong emitters~\cite{StatsJPD2019} which may be presented as rf breakdowns. 3) At higher fields, $sp^{2}$--poor (N)UNCD sites dominate the emission as $sp^2$--rich locations on (N)UNCD surface are consumed as conditioning proceeds, leading to low $\beta$ and high $E_{turn-on}$. This speculation is consistent with the monotonically decreasing $\beta$ and the increasing $E_{turn-on}$ observed in the experiment. It is also supported by the observation of dramatic emission property changes during the conditioning process with a large number of rf breakdowns versus nearly unchanged emission properties in the longevity measurement with only one rf breakdown detected.
	
	It should be noted that in the context of the obtained results and the proposed emission mechanism, the near surface region always has a certain amount of  accumulated charge that makes the charge-field relation Fowler-Nordheim-like, with no observable saturation effects~\cite{ChubenkoJAP2019} due to low duty cycle of the ACT beamline. Therefore, the application of the classical Fowler-Nordheim law is plausible: it simplifies analysis and makes comparison between different geometries and material modifications reasonably good. Strictly speaking, the Stratton-Baskin-Lvov-Fursey formalism should be considered for semi-conductor/semi-metal material analysis and modelling: this is however an exceptionally complex problem that is out of the scope of the presented study.	
	
	\section{\label{cha6}Summary and future studies}
	This study systematically benchmarked the field emission properties of a planar nitrogen-incorporated ultrananocrystalline diamond during the rf conditioning process where the macroscopic field was pushed from $\sim$8~MV/m to $\sim$42~MV/m. The cathode reached a maximum charge of 15~nC and an average emission current of 6~mA during a 2.5~$\mu$s emission period. The charge dropped by only $\sim$4~\% during a 4-hour longevity measurement at 42~MV/m which accumulated $\sim 3 \times 10^{4}$ rf pulses or $\sim 1 \times 10^{8}$ rf cycles. The high resolution field emission images revealed a large number of field emitters on the cathode with a density over 100/cm$^{2}$. This study demonstrates the good potential of (N)UNCD cathodes for application in FEC-based injectors. The observed conditioning effects can all be reasonably explained using the band structure of the material, which, in turn, provides insight to guide further material engineering for improved cathode performance. Future work includes experiments to study the emittance of (N)UNCD field emission cathodes and the design of electron sources based on the parameter space reported in this manuscript.
	
	\begin{acknowledgments}
		The work at AWA is funded through the U.S. Department of Energy Office of Science under Contract No. DE-AC02-06CH11357. The work by Mitchell Schneider is supported by the US Department of Energy, Office of Science, High Energy Physics under Cooperative Agreement award No. DE-SC0018362 and Michigan State University. The work by Gongxiaohui Chen is supported by NSF grant No. NSF-1739150, and NSF-1535676. The work by Sergey Baryshev is funded from the College of Engineering, Michigan State University, under the Global Impact Initiative. Use of the Center for Nanoscale Materials, an Office of Science user facility, was supported by the U.S. Department of Energy, Office of Science, Office of Basic Energy Sciences, under Contract No. DE-AC02-06CH11357. We would like to thank Dr. Evgenya Simakov at Los Alamos National Laboratory for providing the molybdenum cathode.
	\end{acknowledgments}
	
	\bibliography{UNCD_ref}
	
\end{document}